\def\year{2022}\relax
\documentclass[letterpaper]{article} 
\usepackage{aaai22}  
\usepackage{times}  
\usepackage{helvet}  
\usepackage{courier}  
\usepackage[hyphens]{url}  
\usepackage{graphicx} 
\usepackage{natbib}  
\usepackage{caption} 
\DeclareCaptionStyle{ruled}{labelfont=normalfont,labelsep=colon,strut=off} 
\usepackage{pdfpages} 
\usepackage{algorithm}
\usepackage{algorithmic}

\usepackage{graphicx}
\usepackage{amsmath}
\usepackage{caption}
\usepackage{subcaption}
\usepackage{nameref}
\usepackage{inconsolata}
\usepackage{enumitem}
\usepackage{newfloat}
\usepackage{listings}
\floatstyle{ruled}
\newfloat{listing}{tb}{lst}{}
\floatname{listing}{Listing}
\nocopyright
\title{A Novel Method for Analysing Racial Bias: Collection of Person Level References}
\author {
    Muhammed Yusuf Kocyigit \textsuperscript{\rm 1} \hspace{2pt}
    Anietie Andy \textsuperscript{\rm 2} \hspace{2pt}
    Derry Wijaya \textsuperscript{\rm 1}
}
\affiliations {
    \textsuperscript{\rm 1} Boston University \textsuperscript{\rm 2} University of  Pennsylvania\\
}

\begin{document}
\maketitle
\begin{abstract}
Long-term exposure to biased content in literature or media can significantly influence people's perceptions of reality, leading to the development of implicit biases that are difficult to detect and address \cite{gerbner1998cultivation}. In this study, we propose a novel method to analyze the differences in representation between two groups and use it examine the representation of African Americans and White Americans in books between 1850 to 2000 with the Google Books dataset \cite{goldberg-orwant-2013-dataset}. By developing better tools to understand differences in representation, we aim to contribute to the ongoing efforts to recognize and mitigate biases. To improve upon the more common phrase-based (men, women, white, black, etc) methods to differentiate context \cite{Tripodi2019TracingAL, Lucy2022DiscoveringDI}, we propose collecting a comprehensive list of historically significant figures and using their names to select relevant context. This novel approach offers a more accurate and nuanced method for detecting implicit biases through reducing the risk of selection bias. We create group representations for each decade and analyze them in an aligned semantic space \cite{hamilton-etal-2016-diachronic}. We further support our results by assessing the time-adjusted toxicity \cite{Bassignana2018HurtlexAM} in the context for each group and identifying the semantic axes \cite{Lucy2022DiscoveringDI} that exhibit the most significant differences between the groups across decades. We support our method by showing that our proposed method can capture known socio-political changes accurately and our findings indicate that while the relative number of African American names mentioned in books have increased over time, the context surrounding them remains more toxic than white Americans.

\end{abstract}

\section{Introduction}
In the last two hundred years, perhaps with the exception of the last decade, books have been one of the most important medium for the dissemination and perpetuation of knowledge and culture across generations. 

Analyzing the content of books through time resembles examining the rings in the cross section of a tree. It can tell us about the past and help us understand the present. By studying how different groups of people are portrayed in books throughout history, we aim to better understand the ways in which racial bias has manifested and developed through time. 

\begin{center}
\begin{figure}[ht]
    \centering
    \includegraphics[scale=0.55]{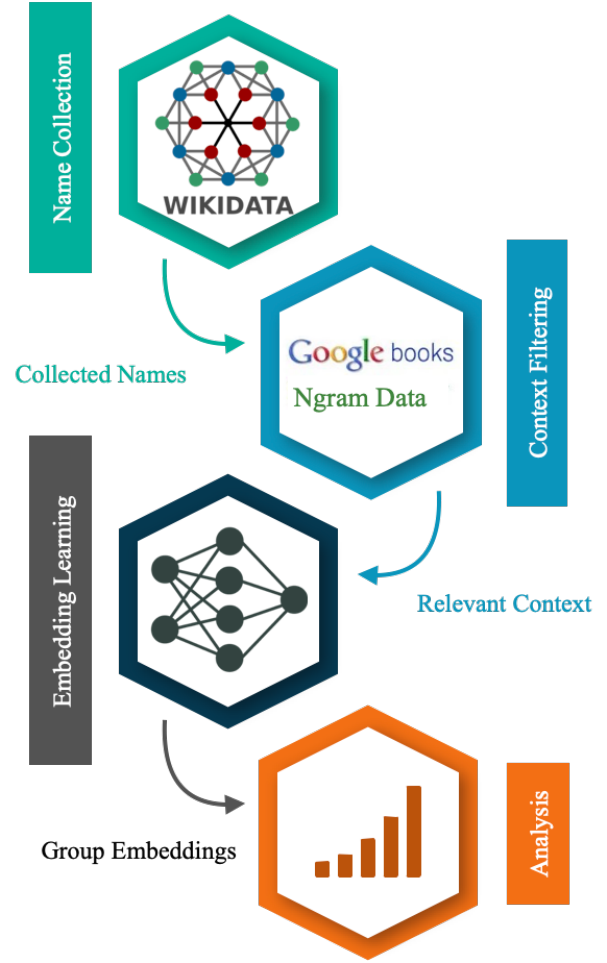}
    \caption{High-level summary of the pipeline of our method. We collect names and retrieve context for each group using these names from the Google Ngrams Data. Using the context, we learn contrastive group embeddings for each group per decade and analyze the embeddings.}
    \label{fig:pipeline}
    \vspace{-10pt}
\end{figure}
\end{center}

Recent advances in text representation and content analysis has enabled us to process large corpora and create meaningful representations that we can analyze, enabling large scale, longitudinal text based studies. In this work we aim to integrate and enhance a diverse array of methods to create a toolbox for analyzing such content. 

Our method\footnote{We will release the data and code to encourage further development and application of our method. } as summarized in Figure \ref{fig:pipeline} consists of 4 stages. First we collect a large sample of individuals that represent the groups we will analyze. In our case we want to examine the context surrounding African-Americans and White Americans in books published in the United States between 1850-2000. So we collect the name and ethnicity of every person who has lived in the United States from Wikipedia. Then we take the Google Books dataset and filter the n-grams that contain any of the names that we have collected. We use these n-grams to train embeddings that represent each ethnic group and finally use these embeddings to study the differences in the context in which each group is talked about and how it has changed over 15 decades. The longitudinal nature is a vital aspect of this study as it allows us to study biases that are cultivated through extended exposure. These biases can affect the base frame that people ground their notion of normalcy in. Thus,  understanding them presents a critical issue since gaining awareness of one's sense of normalcy is a challenging task without the aid of explicit external information and intentional inquiry. 

While similar work exists at the intersection of Natural Language Processing and social sciences, our work has some important differences. Researchers have used word representations in combination with other text processing tools to represent changes in the meanings of words \cite{martinc-etal-2020-leveraging}, analyze online communities to detect gender bias \cite{Lucy2022DiscoveringDI}, analyze literary work to detect racial bias \cite{Tripodi2019TracingAL}, or quantify ethnic stereotypes in the news \cite{sorato-etal-2021-using}. However, these methods predominantly use referring words (woman, man, African American, Jewish, etc.) or gendered terms (he, she, lady, gentleman, etc.) to differentiate each group's context. 

Instead of referring phrases, in this work, we directly employ a list of names of African American and White American figures (politician, athlete, musician, and social leader categories) for each decade from Wikipedia. While any representation method has downsides, we propose that this personal-based representation may better capture implicit biases such as coded language (covert expression of racism that avoids any mention of race)\footnote{\url{https://www.r2hub.org/library/overt-and-covert-racism}} particularly as names do not have explicit references to a larger group. Although individuals may possess knowledge of the race of the person they are writing about, the statements they make are not inherently generalizations. As a result, authors may express their implicit biases openly in their writing. Consequently, if we analyze a substantial corpus of text and find that these instances of bias occur frequently, we can identify a systematic pattern of implicit bias. Hence, we use how individuals from a racial group are presented to analyze biases for the general group. In addition, investigating personal-based racial bias holds significance as studies have shown its association with more detrimental effects on mental health than group-level bias \cite{hagiwara2016differential}. 

\begin{center}
\begin{figure}[h]
    \centering
    \includegraphics[scale=0.28]{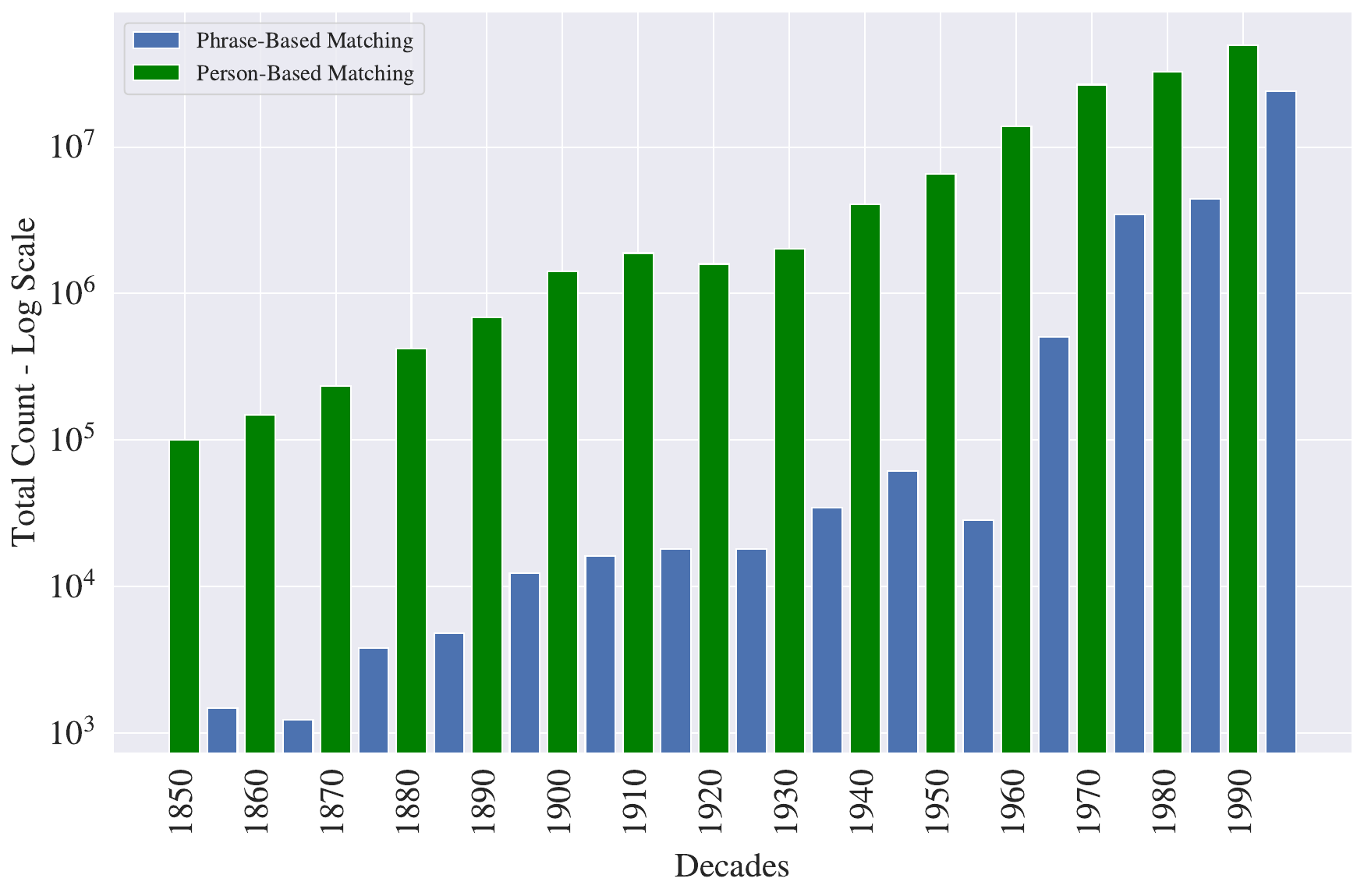}
    \caption{Total number of n-grams matched in Phrase-based filtering of n-grams (Black American$||$African American) compared to Person-based filtering. Note that the y-axis is in log scale so there is almost a 100x difference between the two methods in the scale of data that can be used for analysis. In addition, the data obtained before 1950s is fairly small with phrase-based methods.}
    \label{fig:filter_statistics}
\end{figure}
\end{center}

Additionally our proposed method gets around the problem of selecting referring phrases. Selecting unbiased, phrases that will return a large enough context for analysis becomes a challenge, if not a roadblock since social-context itself is ever changing and the tools for referring change as well. We present the number of words returned from the complete Google n-grams data with the referring phrase- and person-based filtering methods for 15 decades in Figure \ref{fig:filter_statistics}. The number of n-grams containing referring phrases for the decades before 1970 is around 10K, which is too few considering these n-grams contain a lot of stop words and part-of-speech tags.

Similar to \cite{Lucy2022DiscoveringDI} we use learned group representations as our unit of analysis. Unlike previous work, we propose using a contrastive objective to learn the representations to better capture implicit biases since the absence of a positive context should be just as strong a signal toward bias as the existence of a negative one. Due to its subtle language, implicit biases may be hard to identify when doing a one-sided analysis alone but may be visible when we look at the differences in the context used when referring to people from different racial groups. Contrastive learning allows us to represent positive and negative distributional differences in the context of two groups. 

Additionally, we enhance our analysis by analyzing multiple aspects of the text surrounding each group. We look at the frequency of toxic words in each decade per racial group. We also employ semantic axes \cite{Lucy2022DiscoveringDI} to analyze the difference in the embeddings of the racial groups with respect to a variety of semantic axes and how this changes over the 15 decades. 

\textbf{In summary, our contributions in this work are:}
\begin{itemize}[noitemsep,topsep=0pt]
    \item We propose a novel method of analyzing bias between two groups with person-based filtering that is based on the names of people in a group instead of referring-phrases or group terms.
    \item We use contrastive learning to capture both the presence and absence of relevant information to better represent the difference in context. 
    \item We conduct, to the best of our knowledge, the first large-scale longitudinal study of the bias between African-Americans and White Americans in books published in the United States between 1850-2000.
    \item We orchestrate multiple analysis methods and integrate them with our representation learning method, showing easy integration of our representation learning method with existing literature. In our integration, we also take into account the semantic changes that occur through long time periods.
\end{itemize}

\section{Related Works}

Two primary aspects distinguish works in this area from each other: how the context is selected for each group/community and how the trained embeddings on this chosen text are used. Below we will discuss prior works based on how they use their learned embeddings as well as how they differentiate the context for each group. We use the phrase "context" to refer to words that co-occur with a specific group/community in the corpus. These words are utilized to train the group embeddings and to analyze the portrayal of this group.

\subsection{Word Embeddings Usage}

Word embeddings have been used for detecting larger-scale semantic and cultural shifts. \citet{martinc-etal-2020-leveraging} focused on identifying short-term cultural changes in public discourse.  \citet{doi:10.1177/0003122419877135} analyzed cultural variations in a more extended period, while \citet{hamilton-etal-2016-diachronic, wijaya2011understanding} looked into understanding semantic shifts and how the meaning of words have changed over long periods of time. 

Previous works have also used word embeddings to analyze social dynamics towards or within a specific community. \citet{Farrell2020OnTU} used word embeddings to identify novel jargon generated by the online community \textit{manosphere}. \citet{Tripodi2019TracingAL} tracked antisemitic sentiment over 12 decades in documents they acquired from \textit{Bibliothèque Nationale de France} using words that refer to Jewish People to extract relevant text and generate a semantic space. \citet{Xu_2019} analyzed movie scripts and books to identify gender representations in narratives. \citet{Lucy2022DiscoveringDI} used semantic axes, a bipolar collection of adjectives that represent an antonymous relationship \cite{An2018SemAxisAL} and gendered words to analyze how behavior towards women has changed in different stages of the online \textit{manosphere} community. They show how clustered terminology coincides with the opening of various subreddits and analyze these clusters on different semantic axes. \citet{Garg_2018} used word embeddings in a larger-scale work to analyze ethnic and gender stereotypes as well as how these stereotypes changed over time. Additionally, they combined the findings with demographic and occupation data to offer a more coherent narrative. 

\begin{center}
\begin{figure*}[h]
  \centering
  \begin{subfigure}[b]{0.49\textwidth}
    \includegraphics[width=\linewidth]{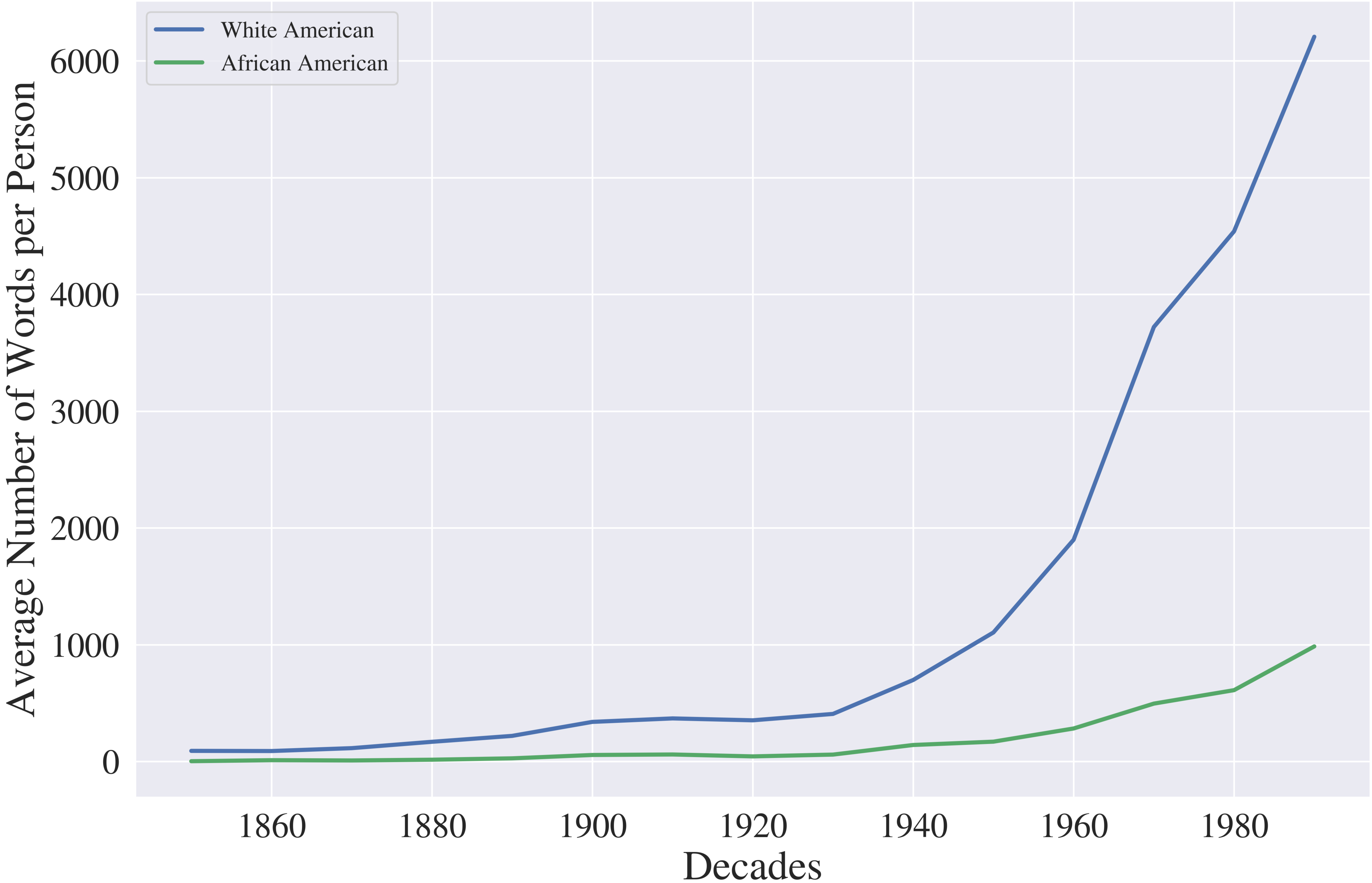}
    \caption{Average Number of Context Words per Person per Decade}
    \label{fig:subplot1}
  \end{subfigure}
  \hfill
  \begin{subfigure}[b]{0.49\textwidth}
    \includegraphics[width=\linewidth]{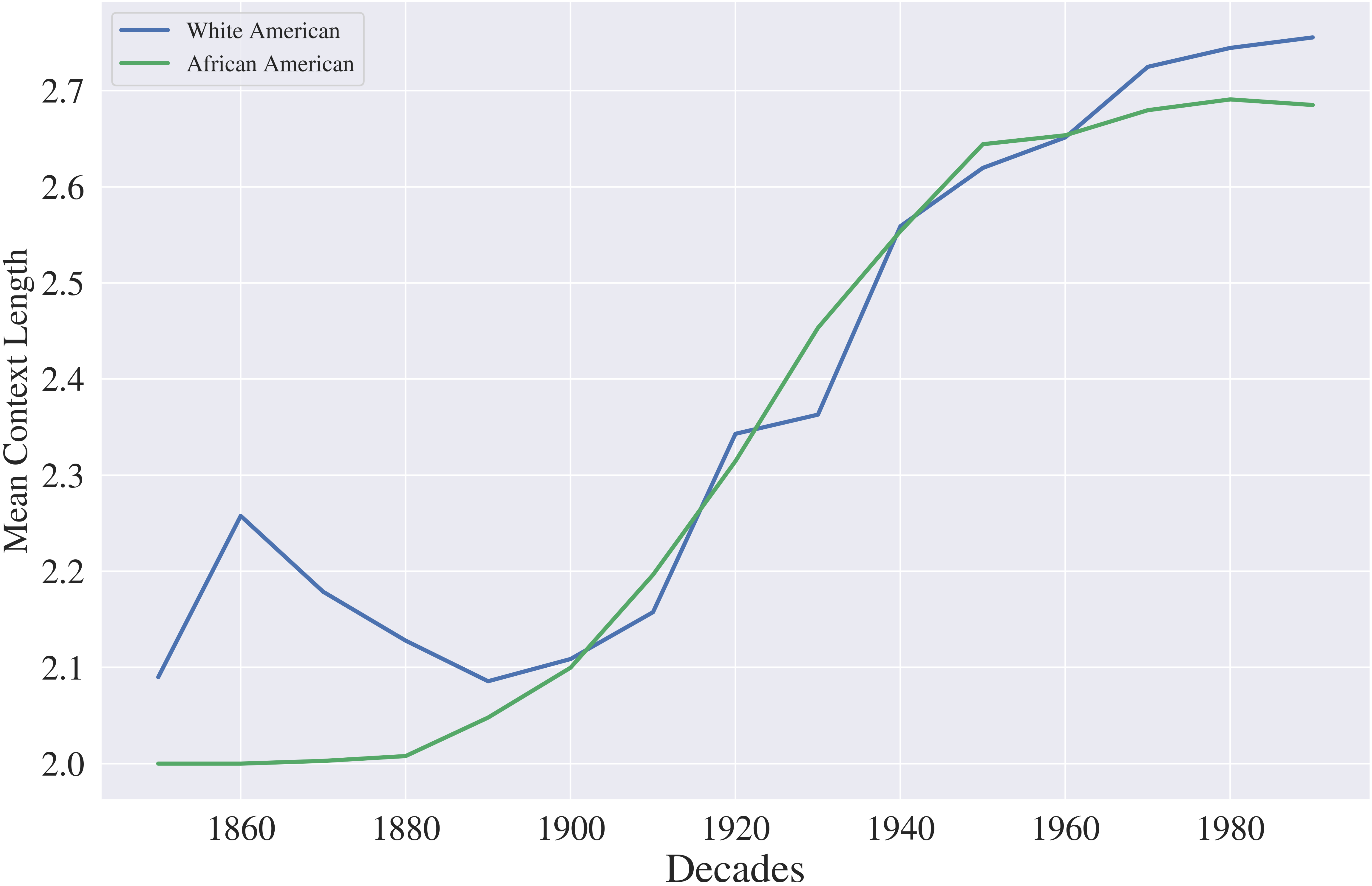}
    \caption{Average Context Length per N-gram per Decade}
    \label{fig:subplot2}
  \end{subfigure}
  \caption{The context length, if too different, can cause bias in the learned embeddings of the names because of how the Google Books dataset is structured. If a name consist of two names the analysis would include context words that are two words away from the name. However, if a name has three words we will only have words that are one word away from the name as context. This difference could have caused differences in the semantic meaning captured by the learned embeddings. However, we observe that the average context length for both groups is fairly similar especially in the later years where the number of extracted n-grams is higher.}
  \label{fig:main}
\end{figure*}
\end{center}

\subsection{Context Differentiation}

Researchers use several methods to filter the context of the target group and differentiate it from other groups with which we will compare the target group. \citet{Bolukbasi2016ManIT} used demographically gendered names to identify the context that is specific to one gender. \citet{An2018SemAxisAL} used a curated list of words that referred to Jewish People and filtered the relevant text. \citet{Lucy2022DiscoveringDI} took this core idea one step further and trained a gender classifier on a predetermined list of phrases and tried to find other keywords for filtering relevant context. \citet{doi:10.1177/0003122419877135} searched for keywords in dictionaries and thesauruses and observed that larger keyword sets yield more reliable results. While many of these methods have yielded meaningful results, we suggest that filtering based on referring words or demographically gendered or racial names is an approximation.. We propose using names of actual people in complete form and only extracting context applicable to them. If applied at enough scale, this method will yield more precise measurements of implicit bias compared to previous methods. 

Additionally, when people use referring phrases, the expression is explicitly general; and biases are generally filtered when making general statements about a group. Changes in laws, such as the Civil Rights Act, and social codes have served to suppress overt expressions of group-level bias. However, when talking about individuals, the group context is not explicit. People can have implicit biases toward a group that they project on the majority of individuals from that group that they are unaware of, or filter when making general statements about the group. Hence the collection of context around individual references is a better manifestation of implicit biases. 

\section{Data}
We collect names of significant figures from Wikidata. We use this collection of names to extract the relevant context n-grams from the Google Books dataset. Additionally, we use the 732 semantic axes presented by \citet{An2018SemAxisAL} to analyze the semantic axes with which the two racial groups (African Americans and White Americans) diverge maximally over time. Finally, we use the time-adjusted Hurtlex \cite{Bassignana2018HurtlexAM} toxic words list to determine the toxic context per racial group over time. 

\begin{center}
\begin{figure}[h]
    \centering
    \includegraphics[scale=0.047]{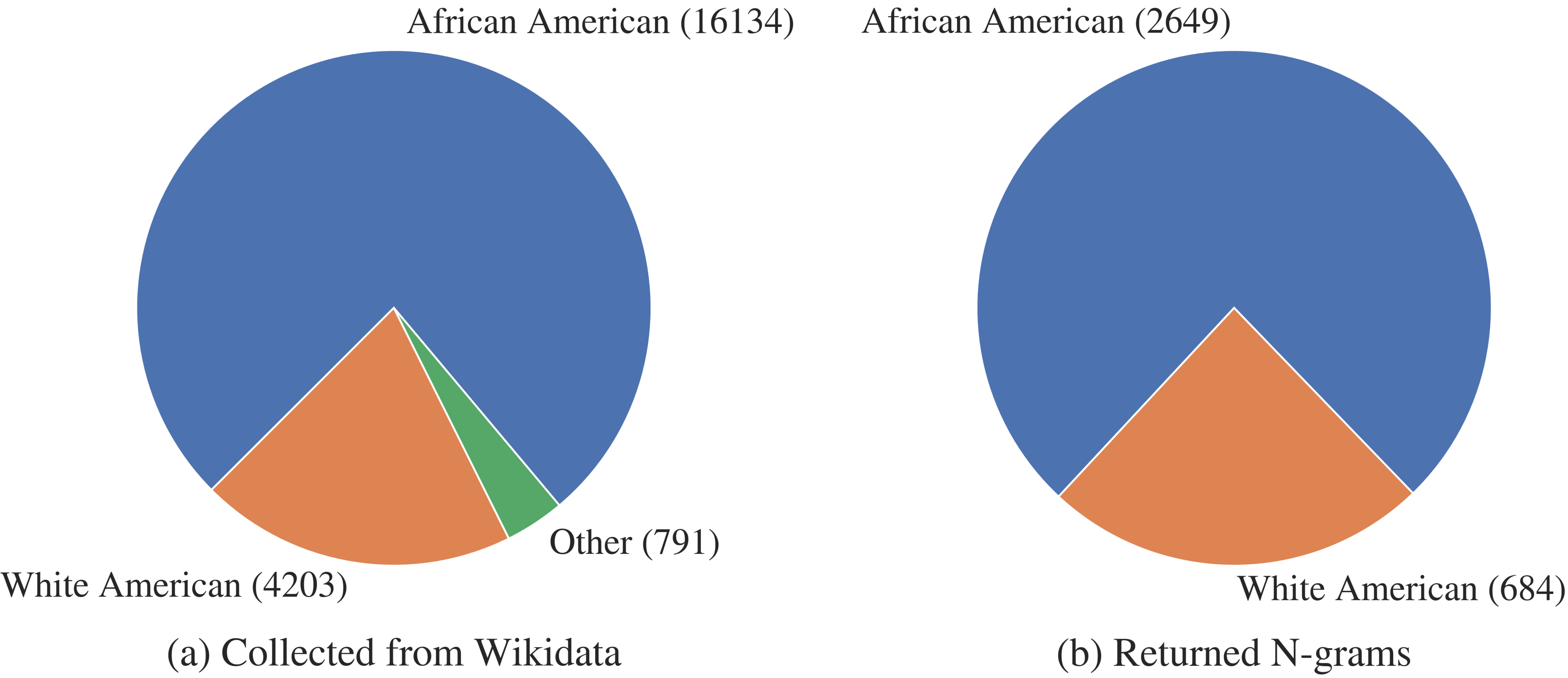}
    \caption{The ethnicity distribution of the names that are extracted from Wikipedia and the distribution of unique people that return any n-gram matches after the n-gram extraction.}
    \label{fig:names}
    \vspace{-10pt}
\end{figure}
\end{center}

\subsection{Wikidata and Extracted Names}

Manually collecting names on a large enough scale requires a considerable amount of time. Instead, we opted for a different approach and searched Wikidata for people who have lived, been a citizen of, or were born in the United States. The details can be found in Figure \ref{fig:wiki} in the Appendix. This query returns around 22K people of which the ethnicity distribution can be found in Figure \ref{fig:names}. We don't exclude or make any explicit filters on this set to get the broadest set of people that we can use in this work. Once we extract n-grams surrounding mentions of these names in Google Books dataset,  
only around 3K of these names return at least one n-gram in the Google Books dataset. We search for complete name matches (first name + last name) and remove the returning n-grams that are before the birth year of the matching person\footnote{We use the assumption that a person must be at least 10 years old before they can be written in books.}. 

\subsection{Google Ngram Filtering}

We work with the 20200217 version of the Google Books Ngram data \cite{goldberg-orwant-2013-dataset} specifically, the 5-gram American English subset. In total we analyze around 140 million 5-grams in 15 decades. The smallest data comes from the earliest decade 1850-1860 with around 100K 5-grams, and the largest sample is from the latest decade 1990-2000 with approximately 50M 5-grams. We upscale each n-gram with its count given in the Ngram data and sample the n-grams for each group to a fixed sample to learn its embedding as described in \nameref{section:method} section. 

Figure \ref{fig:main} presents the average number of context words extracted per person per decade and the average context length per group. We observe that the number of context words for White Americans is significantly larger compared to African Americans in each decade. Thus, while the mention of African American figures has increased over time in books, their mention is still smaller in quantity compared to White Americans. We also observe that there is no significant difference in the mean context length (i.e window size) of the two groups that could introduce an unwanted difference in the semantic meaning captured by their learned  embeddings.      
\subsection{Hurtlex}

To analyze the toxicity towards each group we utilize the toxic words list provided by \citet{Bassignana2018HurtlexAM}. We utilize v1.2 English and the "conservative" level as defined by Hurtlex that contains 3360 toxic words. We use these words as a dictionary-based method and count the frequency of words in the context of African Americans or White Americans that are in this toxic words dictionary. However, long-term studies must consider the semantic shift problem, as words can become more or less toxic over time. We tackle this problem with a hybrid method, utilizing the toxic words dictionary and the semantic axes provided by \citet{An2018SemAxisAL}. We provide more necessary detail on how the time-adjusted toxic words are calculated in the Toxicity subsection under Analysis and Results.

\section{Method}
\subsection{Context Filtering}
Isolating the context is the first step of our method. As we have mentioned, we collect every person who lived, been a citizen of or was born in the United States and use their complete names to identify the n-grams that contain the names. For this stage we go over the 140M 5-grams for each of the 20K names we have collected. The experiments are ran over CPU in parallel and aggregated at a later stage. This step is the most computationally expensive step. The training and analysis steps are less compute intensive since these steps work on a smaller subspace and only learns two vectors. 

The n-grams in the Google books dataset is created by collecting repeating n-grams from many books. So for a sentence to make it into the dataset it has to be repeated in other books as well. The dataset also contains n-grams where words are replaced with their part of speech tags. We ignore all part of speech tags in the filtered n-grams. We also remove numbers and stop words since they don't contribute meaningfully to the final embedding. After extracting the context surrounding the these names, we aim to analyze the representation of individuals within that context. 

The use of public figures to represent collective bias can be a challenge. Since public figures, in general, belong to a very small socioeconomic minority in any large social group, this brings up the question of the reliability of  representations obtained using public figures. However, the main dataset used in this work is the Google Books dataset and  statistically, books are more likely to talk about public figures. Since we are trying to analyze the representation in books, using public figures can be postulated to be the best option.

\subsection{Word Representation Learning}
\label{section:method}
In previous studies, researchers trained word representations for each referring word, used these for their analysis and then aggregated results at the final stage. In this paper, we adopt a different approach by consolidating the group into a single entity and conducting all analyses on that entity. To achieve this, we replace all names in the text with a specialized group tokens (e.g., GRP\_A and GRP\_B) and learn representations for these group tokens. We train these group embeddings using a contrastive loss, which samples positive words from the context of that group and negative words from the context of other group.

 As discussed in previous literature \cite{akyurek2022measuring, akyurek2022challenges}, bias can arise not only in the presence of negative context but also from the lack of positive context. To capture this dynamic, we learn group representations with a contrastive loss.

 Formally, if $C_{self}$ and $C_{other}$ are the set of context words for each group and $f_{self}(w_i)$ is the frequency of word $i$ in the self context from which the positive words will be sampled and $f_{other}(w_i)$ is the frequency of the same word in the other set. We sample positive words from the former with the probability in Equation \ref{eq:positive_sample}, and we sample negative words from the latter with probability in Equation \ref{eq:negative_sample}. Intuitively this means, words that are more frequently found in the other group are less likely to be selected as positive samples and vice versa. Similarly, for negative samples, words that appear more frequently in the other group are more likely to be sampled as negatives and vice versa.

 \begin{equation}
     P(w_i) = \frac{f_{self}(w_i)}{max(f_{other}(w_i), 10^{-5})}
     \label{eq:positive_sample}
 \end{equation}

  \begin{equation}
     P(w_i) = \frac{f_{other}(w_i)}{max(f_{self}(w_i), 10^{-5})}
     \label{eq:negative_sample}
 \end{equation}

For general hyper-parameter settings, the choice of the hyper parameter $10^{-5}$ in sampling probabilities was inspired from \citet{hamilton-etal-2016-diachronic}'s sub-sampling scheme for training word representations. For the data sampling step we use $k=500,000$ with $n=4$ for all analysis, where $n$ is the ratio of negative to positive samples and $k$ is the number of positive samples. 
Using a larger number of negative samples in contrastive learning is a commonly accepted approach and the number of negative samples experimented in this study remains relatively modest and aligns with the suggestions put forward by previous work \cite{contrastive_samples,contrastive_2,contrastive_3}.
The training process consists of a sampling and a training phase. After the positive $X_{pos}$ and negative $X_{neg}$ words are sampled, we train an embedding layer to learn representations for the novel tokens "GRP\_A" and "GRP\_B" with similar dimensions as the pre-trained word embeddings released by \citet{hamilton-etal-2016-diachronic}. The trained network consists of a single embedding layer with no bias or non-linearity and is trained with the pairwise ranking loss defined in Equation \ref{eq:contrastive_loss}. Here, the label $y$ for the words from $X_{pos}$ is $1$, and $-1$ for words from $X_{neg}$. 
$m$ is the margin which is set to 0.5, $x$ is the group representation that needs to be learned and $w$'s are pre-trained word representations per decade from \cite{hamilton-etal-2016-diachronic}.

  \begin{equation}
    L(x,w)=
    \begin{cases}
      1- cos(x, w) & \text{if y=1} \\
      max(0, cos(x, w) - m) & \text{if y=-1}
    \end{cases}
    \label{eq:contrastive_loss}
  \end{equation}

We use the learned representations for the group tokens in three main types of analysis.

\section{Analysis and Results}
In this section, we present results that are obtained using the group embeddings trained with the details provided in the Method Section. We will go over each method of analysis and results here.

\subsection{Correlation over Time}
\begin{center}
\begin{figure*}[ht]
    \centering
    \includegraphics[scale=0.067]{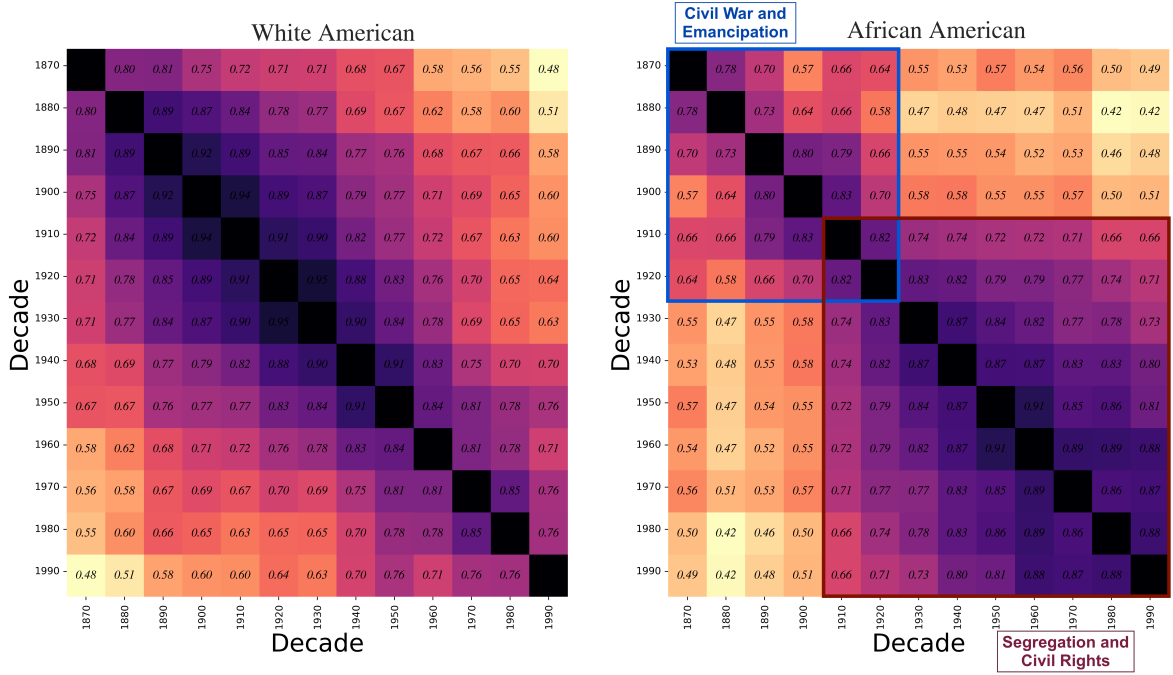}
    \caption{Pearson Correlation between decades of group representations for White Americans (Left) and African Americans (Right). The results indicate that the context surrounding White Americans maintains more consistency over time, no breaks with $p<10^{-4}$ in Kolmogorov–Smirnov(KS) two-sample test, whereas the context related to African Americans displays notable changes 1900s-1910s as tested by KS test. More details in Appendix \ref{appendix_cot}. The observed variations also correlate strongly with historical time periods where there was legal and cultural changes regarding the positions of African Americans in American Society.See [Correlation Over Time subsection of Analysis and Results]}%
    \label{fig:heatmatp}%
\end{figure*}
\end{center}

We first examine the temporal relationships within the context of each group. We calculate the Pearson correlation between each pair of decades for every group as illustrated in Figure \ref{fig:heatmatp}. For example, the top-right cell in Figure \ref{fig:heatmatp} would be the Pearson correlation between the vectors for the African American group embeddings for $1870$ and $1990$. 

The objective of this is two-fold. First, we want to see how the context around each racial group has changed, which decades have exhibited stronger changes compared to others. Second, this we use this as a sanity check mechanism similar to \citet{Garg_2018}. 
Our findings suggest that the context encompassing White Americans exhibits greater stability over time with most decades correlating over 0.6 with any other decade. In contrast, the context pertaining to African Americans demonstrates significant discontinuities. This highlights the potential disparities in the evolution of social contexts surrounding these groups throughout history. 

Additionally we show that the way African Americans are represented in books has drastically changed at the turn of the $20^{th}$ century. This change splits the time frame between 1870-1900 and 1900-2000 corresponding to the Civil War and Emancipation and the Segregation and Civil Rights periods (which we have marked over the heatmap) with a certain delay, which we approximate as 10 years. We also test the significance of the observed difference with the Kolmogorov–Smirnov two-sample test (more details in Appendix: Correlation Over Time) and show that there is a notable change in the period between 1900s-1910s in the African Americans embeddings, indicating that the important social events of the time as well as the civil rights movement had a strong impact on the representation of African Americans in books. 

Literature is naturally influenced by socio-political events but in this work we computationally measure the rate and extent of this influence and show the details of this change as well. Additionally this analysis serves as a verification of the embeddings. The alignment between the results our method has captured and well known historical events is emblematic of the method's capacity to effectively capture and illuminate large-scale trends within literary works. Consequently, our method's ability to reveal outcomes that resonate with our understanding of historical events is a strong signal that the collection of individual representations can be used to measure the representation of the collection of individuals. 

\subsection{Semantic Axes}
Our second approach to analyzing the context involves the use of semantic axes, as introduced by \citet{Lucy2022DiscoveringDI}. Semantic Axes, broadly are collections of adjectives that define a specific relationship. For example an axis could be the "Noble" axis where one (left-hand) pole is defined by words like base, common, and ignoble and the other (right-hand) pole is defined by words like aristocratic, august, and blue. The semantic axis is the vector connecting the mean of the left-hand pole toward the mean of the right-hand pole. Later any word vector can be compared to this semantic axis via cosine distance and analyzed in this direction i.e., nobility in this case. We employ the entire collection of 732 semantic axes presented by \citet{Lucy2022DiscoveringDI}, and represent the poles of these axes with vectors derived from the pre-trained embeddings released by \cite{hamilton-etal-2016-diachronic}. Following \citet{Lucy2022DiscoveringDI}, we exclude any semantic axis that doesn't contain at least three words in both poles.
\vspace{-10pt}

\begin{center}
\begin{figure}[h]
    \centering
    \includegraphics[scale=0.065]{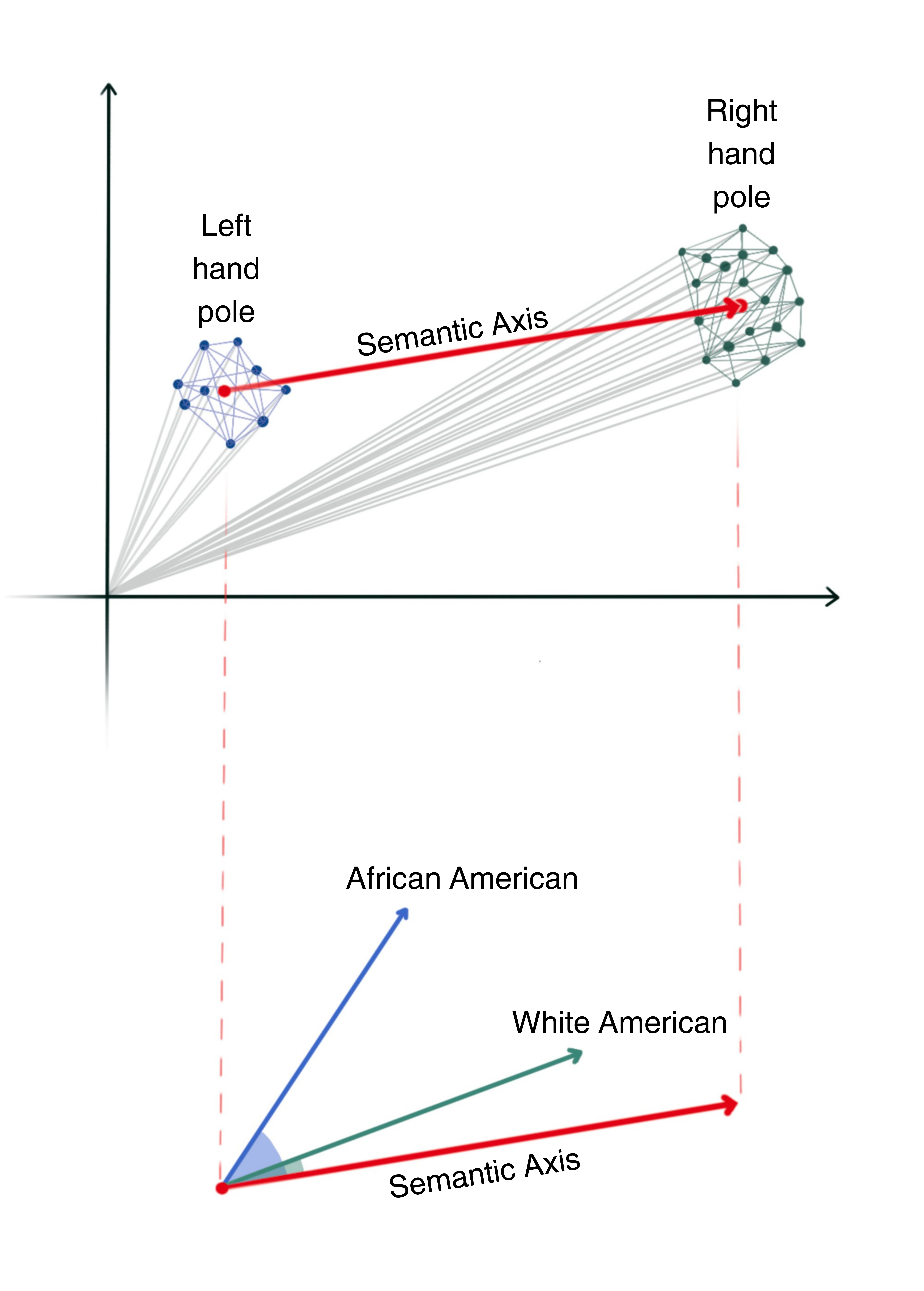}
    \vspace{-8pt}
    \caption{Visualization of the Semantic Axis method presented in \citet{An2018SemAxisAL} and how we utilize this method in this paper. In this representation we display the "noble.a.02" semantic axis from the axes. The right hand pole is contains with words like aristocratic, august, and blue and the left hand pole contains words like base, common, and ignoble. We create the semantic vector and compare the cosine similarity of the representations of African Americans and White Americans from the decade 1850 to this semantic representation and observe that White Americans are much closer to the Semantic Axis.}
    \label{fig:semantic_axes_vis}
    \vspace{-15pt}
\end{figure}
\end{center}

Utilizing the learned group representations, we compute the difference in the cosine similarity between the vector for African American and White American groups to the semantic axes in the same way with \citet{An2018SemAxisAL}. We use the cosine similarity between each group embedding and the semantic vector, then taking the absolute difference between these similarities. In Table \ref{tab:semaxes}, we display these results, with each row corresponding to a decade and showcasing the top-2 semantic axes that exhibit the highest absolute difference between the groups in that decade. The pole words exhibiting greater similarity to each group is placed under its respective column.

\begin{table*}[ht!]
    \tiny
    \centering
    \resizebox{\textwidth}{!}{
    \begin{tabular}{c|r c l|c}
        \multicolumn{1}{c}{ \textbf{Year}} & \multicolumn{1}{c}{\textbf{White American}} &  & \multicolumn{1}{c}{\textbf{African American}} &\multicolumn{1}{c}{ \textbf{Difference} }\\
        \hline
         1850 &\begin{tabular}{r@{}r@{}} \tiny grand, highborn, kingly\\diluted, weak, watery\end{tabular} & \begin{tabular}{c@{}c@{}}  \tiny $\leftrightarrow$  \\ $\leftrightarrow$ \end{tabular} &\begin{tabular}{l@{}l@{}} \tiny untitled, lowly, vulgar\\undiluted, neat, concentrated\end{tabular}&\begin{tabular}{l@{}l@{}}  \tiny 0.28\\0.28\end{tabular}\\
        \hline
        1860 &\begin{tabular}{r@{}r@{}} \tiny superior, dominant, preponderant\\legal, judicial, legitimate\end{tabular} & \begin{tabular}{c@{}c@{}}  \tiny $\leftrightarrow$  \\ $\leftrightarrow$ \end{tabular} &\begin{tabular}{l@{}l@{}} \tiny subordinate, low-level, secondary\\outlawed, illicit, illegitimate\end{tabular}&\begin{tabular}{l@{}l@{}}  \tiny 0.35\\0.25\end{tabular}\\
        \hline
        1870 &\begin{tabular}{r@{}r@{}} \tiny grainy, harsh, granulated\\free, gratuitous, uncompensated\end{tabular} & \begin{tabular}{c@{}c@{}}  \tiny $\leftrightarrow$  \\ $\leftrightarrow$ \end{tabular} &\begin{tabular}{l@{}l@{}} \tiny floury, powdered, tight\\salaried, paying, compensable\end{tabular}&\begin{tabular}{l@{}l@{}}  \tiny 0.2\\0.2\end{tabular}\\
        \hline
        1880 &\begin{tabular}{r@{}r@{}} \tiny standard, grassroots, average\\lost, irrecoverable, unrecoverable\end{tabular} & \begin{tabular}{c@{}c@{}}  \tiny $\leftrightarrow$  \\ $\leftrightarrow$ \end{tabular} &\begin{tabular}{l@{}l@{}} \tiny unusual, red-carpet, special\\recoverable, redeemable, retrievable\end{tabular}&\begin{tabular}{l@{}l@{}}  \tiny 0.27\\0.26\end{tabular}\\
        \hline
        1890 &\begin{tabular}{r@{}r@{}} \tiny confused, alienated, anomic\\eternal, standing, indissoluble\end{tabular} & \begin{tabular}{c@{}c@{}}  \tiny $\leftrightarrow$  \\ $\leftrightarrow$ \end{tabular} &\begin{tabular}{l@{}l@{}} \tiny oriented, headed, familiarized\\jury-rigged, impermanent, evanescent\end{tabular}&\begin{tabular}{l@{}l@{}}  \tiny 0.18\\0.17\end{tabular}\\
        \hline
        1900 &\begin{tabular}{r@{}r@{}} \tiny six-fold, double, sevenfold\\unswerving, straightforward, direct\end{tabular} & \begin{tabular}{c@{}c@{}}  \tiny $\leftrightarrow$  \\ $\leftrightarrow$ \end{tabular} &\begin{tabular}{l@{}l@{}} \tiny one-woman, one-person, lone\\sidelong, diversionary, circuitous\end{tabular}&\begin{tabular}{l@{}l@{}}  \tiny 0.26\\0.22\end{tabular}\\
        \hline
        1910 &\begin{tabular}{r@{}r@{}} \tiny tenderhearted, romantic, affectionate\\exhilarating, renewing, reviving\end{tabular} & \begin{tabular}{c@{}c@{}}  \tiny $\leftrightarrow$  \\ $\leftrightarrow$ \end{tabular} &\begin{tabular}{l@{}l@{}} \tiny unromantic, unloving, uncaring\\debilitating, enervating, draining\end{tabular}&\begin{tabular}{l@{}l@{}}  \tiny 0.24\\0.21\end{tabular}\\
        \hline
        1920 &\begin{tabular}{r@{}r@{}} \tiny pleasurable, dulcet, grateful\\placid, equable, even-tempered\end{tabular} & \begin{tabular}{c@{}c@{}}  \tiny $\leftrightarrow$  \\ $\leftrightarrow$ \end{tabular} &\begin{tabular}{l@{}l@{}} \tiny unhappy, mortifying, acrid\\dyspeptic, bad-tempered, sulky\end{tabular}&\begin{tabular}{l@{}l@{}}  \tiny 0.22\\0.19\end{tabular}\\
        \hline
        1930 &\begin{tabular}{r@{}r@{}} \tiny tractable, vulnerable, pliable\\neat, straight, full-strength\end{tabular} & \begin{tabular}{c@{}c@{}}  \tiny $\leftrightarrow$  \\ $\leftrightarrow$ \end{tabular} &\begin{tabular}{l@{}l@{}} \tiny immune, immunised, immunized\\cut, weakened, washy\end{tabular}&\begin{tabular}{l@{}l@{}}  \tiny 0.17\\0.17\end{tabular}\\
        \hline
        1940 &\begin{tabular}{r@{}r@{}} \tiny undiscovered, unmapped, unbeknownst\\unladylike, unrefined, ungentlemanly\end{tabular} & \begin{tabular}{c@{}c@{}}  \tiny $\leftrightarrow$  \\ $\leftrightarrow$ \end{tabular} &\begin{tabular}{l@{}l@{}} \tiny renowned, noted, celebrated\\twee, ladylike, well-mannered\end{tabular}&\begin{tabular}{l@{}l@{}}  \tiny 0.24\\0.23\end{tabular}\\
        \hline
        1950 &\begin{tabular}{r@{}r@{}} \tiny brutal, direct, straight\\contained, harnessed, possessed\end{tabular} & \begin{tabular}{c@{}c@{}}  \tiny $\leftrightarrow$  \\ $\leftrightarrow$ \end{tabular} &\begin{tabular}{l@{}l@{}} \tiny hearsay, periphrastic, roundabout\\wild, torrential, anarchic\end{tabular}&\begin{tabular}{l@{}l@{}}  \tiny 0.23\\0.2\end{tabular}\\
        \hline
        1960 &\begin{tabular}{r@{}r@{}} \tiny tight, closed, blinking\\reincarnate, material, incarnate\end{tabular} & \begin{tabular}{c@{}c@{}}  \tiny $\leftrightarrow$  \\ $\leftrightarrow$ \end{tabular} &\begin{tabular}{l@{}l@{}} \tiny wide, staring, gaping\\spiritual, incorporeal, disembodied\end{tabular}&\begin{tabular}{l@{}l@{}}  \tiny 0.24\\0.23\end{tabular}\\
        \hline
        1970 &\begin{tabular}{r@{}r@{}} \tiny heavy, soggy, ponderous\\well-connected, engaged, connected\end{tabular} & \begin{tabular}{c@{}c@{}}  \tiny $\leftrightarrow$  \\ $\leftrightarrow$ \end{tabular} &\begin{tabular}{l@{}l@{}} \tiny floaty, buoyant, airy\\separated, obscure, unattached\end{tabular}&\begin{tabular}{l@{}l@{}}  \tiny 0.22\\0.22\end{tabular}\\
        \hline
        1980 &\begin{tabular}{r@{}r@{}} \tiny decided, distinct, certain\\frozen, rooted, motionless\end{tabular} & \begin{tabular}{c@{}c@{}}  \tiny $\leftrightarrow$  \\ $\leftrightarrow$ \end{tabular} &\begin{tabular}{l@{}l@{}} \tiny unfixed, indefinite, coy\\kinetic, automotive, writhing\end{tabular}&\begin{tabular}{l@{}l@{}}  \tiny 0.3\\0.29\end{tabular}\\
        \hline
        1990 &\begin{tabular}{r@{}r@{}} \tiny unsuccessful, stillborn, childless\\upper, high-level, eminent\end{tabular} & \begin{tabular}{c@{}c@{}}  $\leftrightarrow$  \\ $\leftrightarrow$ \end{tabular} &\begin{tabular}{l@{}l@{}} \tiny bountiful, generative, berried\\outclassed, inferior, low-level\end{tabular}&\begin{tabular}{l@{}l@{}}  \tiny 0.3\\0.23\end{tabular}\\
        \hline
    \end{tabular}
    }
    \caption{The top two semantic axes with maximum absolute difference between the two groups per decade. \textbf{An example read:} For 1850, In the semantic axes defined by the adjectives grand $\leftrightarrow$ vulgar, White Americans were closer to the pole defined by the adjective ``grand" and the difference in cosine similarity to this axis between the two groups was 0.28. \textbf{Insight:} We show that before 1940s African American representation is more frequently closer to negative adjectives signaling a long-lasting bias in representation of African American figures in books. See [Semantic Axes subsection of Analysis and Results]}
    \label{tab:semaxes}
\vspace{-5pt}
\end{table*}

In Table \ref{tab:semaxes} we observe that the most significant differences between African Americans and White Americans are in semantic axes related to social status. Between 1850-1990, African Americans were associated with terms such as "lowly", "outlawed" and "inferior" while White Americans were associated with terms such as "grand", "legal" and "eminent". Additionally, if one of the poles of a semantic axis can be identified to have a negative meaning African Americans are consistently related to the negative pole of of that semantic axis with only a few exceptions. 

\subsection{Toxicity}
\label{section:toxicity}

Finally, we investigate the toxicity of the context and how it has evolved over time. To achieve this, we employ a dictionary-based approach using toxic words provided by HurtLex \cite{Bassignana2018HurtlexAM}. However, as words can lose or acquire toxic meanings over time; to account for this, we introduce a novel filtering method. First, we identify semantic axes with the highest average cosine similarity between their poles and the list of toxic words for the latest decade for which we have embeddings, 1990-2000. This process yields a set of axes in which toxic words are predominantly clustered around one of the poles. We select the top 10 semantic axes, and for each decade, we recompute the vectors for these axes and the similarity of toxic words to these axes. We assume that the meaning of a word with respect to toxicity has significantly changed if it switches its side (from being closer to one pole to being closer to the opposite pole) in more than half of these semantic axes. This method is effective in eliminating words that might no longer be toxic; however, we do not search the entire vocabulary for additional toxic words. While extending our approach to include this aspect could be beneficial, it constitutes a separate research direction and is beyond the scope of this paper.
\vspace{-10pt}

\label{section:method_toxicity}
\begin{center}
\begin{figure}[h]
    \centering
    \includegraphics[scale=0.25]{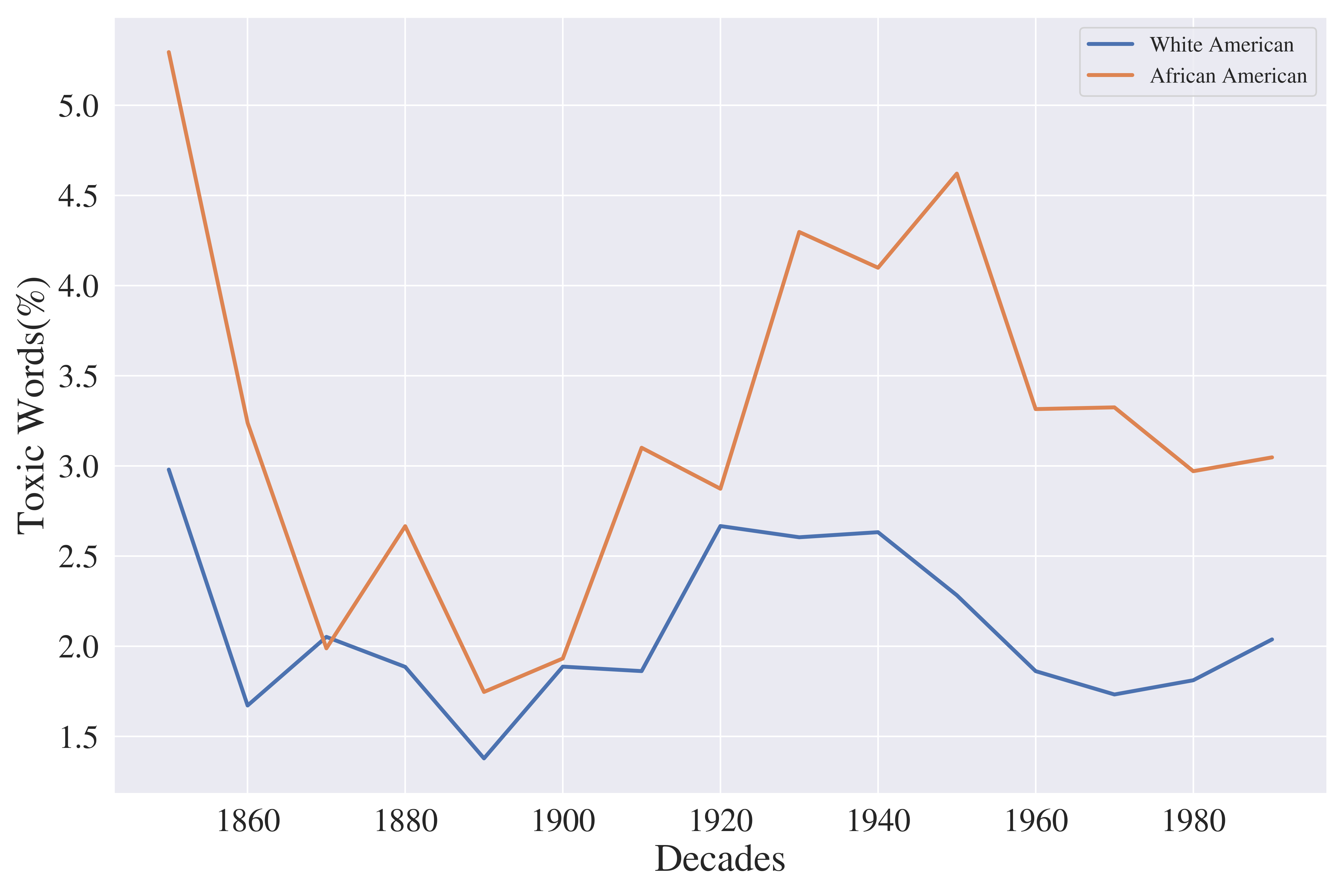}
    \caption{Percentage of Toxic Words for African American and White American figures. We observe that the toxicity for African Americans in books drops after the civil war but rapidly increases after the segregation (1900s) and during the early Civil Rights movement (1950s). While the toxicity has decreased in more recent times, it is still 50\% higher compared to White Americans. See [Toxicity subsection of Analysis and Results]}
    \label{fig:toxicit}
    \vspace{-15pt}
\end{figure}
\end{center}

Our initial hypothesis posited that toxic words would cluster around semantic poles such as "bad" or "wrong." Nonetheless, our experiments revealed that the most extreme poles where toxic words cluster are $\textit{regular} \leftrightarrow \textit{irregular}$, $\textit{educated} \leftrightarrow \textit{ignorant}$, $\textit{superior} \leftrightarrow \textit{inferior}$, and $\textit{normal} \leftrightarrow \textit{abnormal}$, with toxic words clustering around the latter in each case. In the Appendix \ref{sec:appendox_toxicity}, we provide further details on the filtering of toxic words.

Figure \ref{fig:toxicit} shows that the content in books around African Americans is always more toxic compared to White Americans. While the toxicity has dropped after the Civil War, in 1860s and 1870s, this changes and there is an upsurge in toxicity after the 1900s which marks the beginning of the segregation era and continues increasing to the Civil Rights movement (1950s). Finally, in the most recent decades while toxicity has somewhat decreased, African Americans are still exposed to more toxic words and context. 

\section{Conclusion}
We conduct, to the best of our knowledge, the first large scale, longitudinal investigation into books in the United States for investigating racial bias towards African Americans. Our study is also unique in the sense that we use person-names directly, which we crawled from Wikipedia, instead of referring phrases. We also adapt our representation learning method to accommodate both the absence of positive and the presence of negative context to better analyze the bias towards African Americans. 

We show that there exist remarkably higher toxicity, 50\% higher, around African American figures in books. This is problematic since books constitute a large part of our cultural heritage. In Table \ref{tab:semaxes} we illustrate the differences in representation that take into account the learned semantic space. In the presented semantic axes, the difference is significant with African Americans positioned closer to negative adjectives defined by words such as  \textit{lowly}, \textit{obscure}, \textit{inferior}, \textit{subordinate} or \textit{illegitimate}. These results support the claim that African Americans are presented in a more negative and toxic context in books over time.

Additionally, we display the timeline of how African American representation has changed, in which the biggest change has occurred around the beginning of the 20\textsuperscript{th} century. This decade, 1900s, when investigated together with previous results, is also the time where we start observing much higher toxicity, putting additional emphasis that the context around African Americans might have been negatively inclined for a long time. 

Our study provides important insights into the representation of African Americans in literature and written medium over time. The observed discontinuities in the representation of African Americans highlight the ongoing struggle for representation and equality in society and emphasize that while a lot of progress has been made there remains steps to be taken. Finally, our proposed method for analyzing personal-level bias provides a promising approach to understanding the implicit biases present in books, and may serve as a useful tool for addressing racial bias in literature and media.

Our method also enables a new approach to analyzing the differences between group representations without the need for selecting referring phrases. This is especially useful in cases where context is changing, making selecting referring phrases challenging without making assumptions, or selecting unbiased referring phrases is itself a roadblock.

\section{Limitations}
\subsection{Reliability of Word Embeddings}

The representation power and how much word embeddings can model our beliefs is an active point of discussion. This line of research is critical because it allows differentiating under which conditions word representations generate good abstractions to analyze. The critique can be summarized in two main lines: structural and statistical problems. 

Structural problems \cite{Arseniev-Koehler} relate to the Distributional Hypothesis \cite{harrisdist} that current word embeddings are based on. This is criticized by structuralism in multiple aspects. Primarily issues arise when the phrase \textit{a word is known by the company it keeps} is interpreted as \textit{a word \textbf{is} the company it keeps}. The representations of words or groups that we present are not to say that the content represents them precisely as the representations, but they are more closely linked to certain concepts if they are closer in the semantic space. The second important issue we must be aware of is that concepts are not always dichotomous or binary. They can be graded or related in multiple ways. We, however, in line with \citet{An2018SemAxisAL} assume a dichotomous nature to a set of adjectives.

The statistical problems mainly stem from the frequency of different words/concepts. \citet{Valentini2022TheDO} shows that word representations tend to yield higher similarity scores between high-frequency words than between high and low-frequency words. \citet{Joseph2020WhenDW} show that word representations produce similarity measures that better represent human beliefs in with higher frequency words. \citet{Loon_Giorgi_Willer_Eichstaedt_2022} argue that word embedding methods like word2vec cluster high-frequency words together and low-frequency words together. They argue that people have an intuitive positive bias where positive words are more frequent and if a racial/social group(black names in their case) is less frequently represented in the corpus compared to their control group(white names). They are just likely to be related to negative words and this relation would not necessarily be an indication of bias in the corpus. 

Structural problems are related to how we analyze the results and make inferences from them. However, in this paper, we target statistical problems on two levels. First, we look at the distribution of the context words in the primary data where the word vectors were trained and observe that the context words for these groups are not significantly different, and on the frequency of names, we sample equal amounts of (word, context) pairs for both African Americans and White Americans and train the embeddings for these groups from scratch rather than using a pre-trained embedding that would depend on the frequency of names in the initial corpus. 

One specific limitation that we had to work around in this project was the issue of false positive matches. Even though we search for exact name+surname combination in books the resulting n-grams are not necessarily always about the same person that we intent on. While a person with the same name+surname combination likely belongs to the same racial group in our analysis, this might not be the same for other use cases making this a relevant limitation to consider. 

Another relevant limitation is the data size limitation. We have replicated our results with a manually curated name subsets however a smaller subset of n-grams was not a viable option. We have also observed that decades before 1850s return too few samples and the results are not meaningful for those decades. The question of how many n-grams would be enough for a new application is still an open question that we were not able to answer. This question would also extend the scope of this paper beyond a reasonable limit.

\section*{Ethics Statement}
Language and context are not simple things especially considering we are working on a summarization of n-grams instead of sentences. Hence the results and analysis must be evaluated with this in mind. Also the identification of bias can sometimes inadvertently be affected by the researchers' own biases, despite our best efforts to be rigorous in our methodology. A risk lies in potentially over-simplifying or misrepresenting complex socio-cultural dynamics. We have worked to minimize this risk through methodological transparency, iterative analyses, and peer review.

The study inherently involves discussing racially biased language, which could cause distress to some readers, particularly those from communities affected by such biases. Our presentation aims to be sensitive, focusing on systemic patterns rather than individuals or specific works.

Lastly, there is a potential risk of misuse of our findings. The purpose of our study is to raise awareness and prompt action towards eliminating bias, but we are aware that the same information could be used to support biased arguments or perpetuate harmful stereotypes even though the content of books are not a representation of reality but the representation of the perception of the authors which could be source of bias in some cases.

On the positive side, our research contributes to a better understanding of systemic racial biases in historical and cultural contexts. This deeper awareness can help us understand what we perceive as normal or what the next generation will perceive as normal putting a external reality check on the culture we inherit without knowing. 

As a part of our ethical commitment, we will make our data and code public to invite critique, replication, and improvement. Our research is not an end but a beginning, providing a stepping stone for further inquiries into racial bias.
\bibliography{aaai22}


\appendix

\section{Appendix: Toxicity Measurement}
\label{sec:appendox_toxicity}
For toxicity measurements we use a filtering method to remove words that might lose their toxic meaning over time. We use the conservative subset from Hurtlex \cite{Bassignana2018HurtlexAM} and we include all categories off toxic words in our analysis. In Table \ref{tab:removed} we see the words that are removed from the toxic words list. A few interesting observations are words relating to prostitution are generally removed in the early decades. This could be that while they could still have taboo, they had a big enough difference in how they relate to other words in that time that they were filtered out. Another interesting observation is that the word \textit{fascist} was filtered out during the decades 1930 and 1940. This is before World War II where a lot of nationalist sentiment was becoming mainstream in the World which could also explain this change in the meaning.

\begin{table*}[]
    \centering
    \resizebox{\textwidth}{!}{
    \begin{tabular}{c|l}
    \textbf{Decade} & \textbf{Removed Words}\\
    \hline
         1850 & know, demanding, parochial, assumption, scribe, parochial, violation, deserter, con, demanding, derogatory, mess\\
1860 & know, mean, parochial, assumption, parochial, mean, violation, hooker, hooker, con, criminal, derogatory, bed, mean\\
1870 & demanding, hooker, hooker, demanding, fatty, criminal\\
1880 & john, vagina, problem, violation, hooker, deserter, hooker, fatty, criminal, do\\
1890 & know, john, people, problem, violation, story, fatty, operator, assault, story, academic, academic, bed, belligerent, refuse\\
1900 & vagina, demanding, insular, violation, hooker, hooker, demanding, insular, do, bed\\
1910 & problem, insular, violation, fatty, insular, adolescent, operator, academic, academic\\
1920 & know, people, problem, assumption, violation, operator, derogatory, bed, scrapped, scholastic\\
1930 & mean, problem, fascist, assumption, mean, violation, operator, academic, academic, mean, scrapped, scholastic\\
1940 & problem, fascist, violation, bed, scholastic\\
1950 & parochial, assumption, parochial, violation, fatty, juvenile, operator, scholastic\\
1960 & mark, problem, mark, assumption, mark, mark, violation, mark, operator, mark\\
1970 & know, assumption, violation, operator, criminal, academic, academic\\
1980 & mean, problem, assumption, mean, violation, low, academic, academic, eff, mean, scholastic\\
1990 & rookie, people, assumption, cop, adolescent, operator, criminal, outlawed, scholastic\\
    \end{tabular}
    }
    \caption{Caption}
    \label{tab:removed}
\end{table*}

\section{Appendix: Correlation Over Time}
\label{appendix_cot}

We plot the correlation over-time to get a general picture of the embedding space in Figure \ref{fig:heatmatp}. To support our observations we also conduct a Kolmogorov–Smirnov test. We use the two-sided test where the null-hypothesis is that the two empirical distributions are the same. We simply take two columns from our heatmap, ignore the rows where either of the entries are 1 and take the difference and then the absolute between the two lists. The resulting list consitutes our samples from the first distribution for our KS test. The samples from the second distribution is simply the same list for every other transition in our heatmap appended together, since the KS test is not dependent on the number of samples we can run the test for each transition. 

Below in Table \ref{tab:KS_aa} and \ref{tab:KS_aw} the results for the KS test are given. The test simply tells if the two empirical distributions are likely to be from the same distribution. We observe that there are two cases where we can reject the null hypothesis relatively safely. One is for the White Americans heatmap between the years 1920-1930 and the second is for the African American heatmap between 1900-1910. For the first one we observe that the average similarity is well above the average similarity of samples from distribution 2 signaling that the null hypothesis was rejected not because the difference in this transition is big but the contrary. To our point, we observe that for the latter of the two cases the average similarity is much smaller.

\begin{table*}[]
    \centering
    \resizebox{\textwidth}{!}{
    \begin{tabular}{c|c|c|c|c}
    \textbf{Transition Interval} & \textbf{Test Statistic} & \textbf{p value} & \textbf{Mean Distance in Interval} & \textbf{Mean Distance in Rest}\\
    \hline
       1870-1880 & 0.446 & 0.0256751 & 0.06 & 0.049\\
1880-1890 & 0.446 & 0.0256751 & 0.076 & 0.048\\
1890-1900 & 0.347 & 0.143854 & 0.046 & 0.051\\
1900-1910 & 0.744 & 4.2e-06 & 0.123 & 0.044\\
1910-1920 & 0.512 & 0.0061405 & 0.078 & 0.048\\
1920-1930 & 0.248 & 0.503801 & 0.064 & 0.049\\
1930-1940 & 0.289 & 0.3169132 & 0.027 & 0.052\\
1940-1950 & 0.496 & 0.0089816 & 0.025 & 0.053\\
1950-1960 & 0.397 & 0.06457 & 0.019 & 0.053\\
1960-1970 & 0.347 & 0.143854 & 0.028 & 0.052\\
1970-1980 & 0.24 & 0.5468734 & 0.036 & 0.051\\
1980-1990 & 0.421 & 0.041339 & 0.02 & 0.053\\
    \end{tabular}
    }
    \caption{Kolmogorov–Smirnov two-sample test for the African American Embeddings in Table \ref{fig:heatmatp}}
    \label{tab:KS_aa}
\end{table*}

\begin{table*}[]
    \centering
    \resizebox{\textwidth}{!}{
    \begin{tabular}{c|c|c|c|c}
    \textbf{Transition Interval} & \textbf{Test Statistic} & \textbf{p value} & \textbf{Mean Distance in Interval} & \textbf{Mean Distance in Rest}\\
    \hline
       1870-1880 & 0.215 & 0.6808459 & 0.055 & 0.04\\
1880-1890 & 0.562 & 0.0017818 & 0.064 & 0.039\\
1890-1900 & 0.331 & 0.1831865 & 0.028 & 0.042\\
1900-1910 & 0.57 & 0.0014283 & 0.02 & 0.043\\
1910-1920 & 0.24 & 0.5468734 & 0.038 & 0.041\\
1920-1930 & 0.694 & 2.88e-05 & 0.011 & 0.044\\
1930-1940 & 0.397 & 0.06457 & 0.061 & 0.039\\
1940-1950 & 0.14 & 0.9767882 & 0.039 & 0.041\\
1950-1960 & 0.463 & 0.0183634 & 0.054 & 0.04\\
1960-1970 & 0.14 & 0.9767882 & 0.047 & 0.041\\
1970-1980 & 0.355 & 0.1268977 & 0.029 & 0.042\\
1980-1990 & 0.314 & 0.2303857 & 0.048 & 0.04\\
    \end{tabular}
    }
    \caption{Kolmogorov–Smirnov two-sample test for the White American Embeddings in Table \ref{fig:heatmatp}}
    \label{tab:KS_aw}
\end{table*}

\section{Appendix: Wikidata}
\begin{figure}[h]
    \centering
    \begin{verbatim}
    """
SELECT DISTINCT ?item ?itemLabel  
?occupation ?ethnic
?ethnicLabel ?residence 
?place_of_birth ?citizen ?dob WHERE {
    {
        ?item p:P19 ?place_of_birth.
        ?place_of_birth ps:P19 wd:Q30.
    }
    UNION
    {
        ?item p:P551 ?residence.
        ?residence ps:P551 wd:Q30.
    }
    UNION
    {
        ?item p:P27 ?citizen.
        ?citizen ps:P27 wd:Q30.
    }
    ?item wdt:P106 ?occupation.
    OPTIONAL
    {
      ?item wdt:P172 ?ethnic.
    }
    ?item wdt:P569 ?dob.
      SERVICE wikibase:label \\
    { bd:serviceParam \\
    wikibase:language "en".}
    }
"""
\end{verbatim}
    \caption{Wikidata query for extracting individuals that are born in, residents of or citizens of the United States.}
    \label{fig:wiki}
\end{figure}

We use the the query in Figure \ref{fig:wiki} to select the set of individuals that were born in, residents of and citizens of the United States of America. The query takes the ethnic label manually. The ethnic label returns classes that are much more fine-grained then we aim for in this study so we manually create a dictionary to map each sub-group into our main categories presented in Figure \ref{fig:names} as African American, White American and Others. For African American the main rule was that the origin country for the ethnicity would be in the African Continent. We have also classified each European American(for example Italian American, Irish American etc.) ethnicity into White Americans. 

Finally we manually label individuals that are are significant figures that don't contain the ethnicity label in their Wikipedia page. This was more prevalent in White Americans.

\section{Appendix: Hyperparameter Sensitivity}
We conduct our experiments with an array of hyper-parameters to observe the effects of the sampling size and the negative sampling ration and observe that the results are relatively stable. Below are the results for k=[500K, 1M] and n=[1,4,10,20]

\begin{figure*}
    \centering
    \includegraphics[scale=0.067]{hyperparameter/k500K_n1.pdf}
    \caption{Correlation heatmap for k=500K, n=1. Left-White Americans, Right-African American}
    \includegraphics[scale=0.067]{hyperparameter/k500K_n4.pdf}
    \caption{Correlation heatmap for k=500K, n=4. Left-White Americans, Right-African American}
    \includegraphics[scale=0.067]{hyperparameter/k500K_n10.pdf}
    \caption{Correlation heatmap for k=500K, n=10. Left-White Americans, Right-African American}
    \label{fig:hp1}
\end{figure*}

\begin{figure*}
    \centering
    \includegraphics[scale=0.067]{hyperparameter/k500K_n20.pdf}
    \caption{Correlation heatmap for k=500K, n=20. Left-White Americans, Right-African American}
    \includegraphics[scale=0.067]{hyperparameter/k1M_n1.pdf}
    \caption{Correlation heatmap for k=1M, n=1. Left-White Americans, Right-African American}
    \includegraphics[scale=0.067]{hyperparameter/k1M_n4.pdf}
    \caption{Correlation heatmap for k=1M, n=4. Left-White Americans, Right-African American}
    \label{fig:hp2}
\end{figure*}

\begin{figure*}
    \centering
    \includegraphics[scale=0.067]{hyperparameter/k1M_n10.pdf}
    \caption{Correlation heatmap for k=1M, n=10. Left-White Americans, Right-African American}
    \includegraphics[scale=0.067]{hyperparameter/k1M_n20.pdf}
    \caption{Correlation heatmap for k=1M, n=120. Left-White Americans, Right-African American}
    \label{fig:hp3}
\end{figure*}

\end{document}